# Multi-level Loop-less Algorithm for Multi-set Permutations

Tadao Takaoka
Department of Computer Science, University of Canterbury
Christchurch, New Zealand
Email: tad@cosc.canterbury.ac.nz

Abstract: We present an algorithm that generates multiset permutations in O(1) time for each permutation, that is, by a loop-less algorithm with O(n) extra memory requirement. There already exist several such algorithms that generate multiset permutations in various orders. For multiset permutations, we combine two loop-less algorithms that are designed in the same principle of tree traversal. Our order of generation is different from any existing order, and the algorithm is simpler and faster than the previous ones. We also apply the new algorithm to parking functions.

## 1. Introduction.

O(1) time generation for combinatorial objects, such as permutations and combinations, is known also as loop-less algorithms. That is, given the current object, we generate the next object loop-lessly, e.g., with a finite number of statements. This is also called combinatorial Gray code, since the idea is a generalization of binary Gray code to more general combinatorial objects. Ehrich [5] was the first to investigate this topic. Since then many loop-less algorithms were invented for various objects, such as permutations, combinations, parenthesis strings, etc. To list up just a few, see Bitner, et al. [1], Ehrlich [5], Lehmer [12], Eades and McKay [4], Chase [3] for combinations, Johnson [7] and Heap [6] for permutations, Mikawa and Takaoka [13], Vajnovszki [19], and Walsh [21] for parenthesis strings. We focus on the recent topic of multiset permutations in this paper.

There have been several algorithms that generate multiset permutations in O(1) time per permutation. Korsh and Lipschutz (1997) [10] was the first to investigate this problem. This algorithm achieves O(1) time, but uses a linked list for a container of permutations. Thus we can go from (1, ..., 1, 2, ..., 2) to (2, ..., 2, 1, ..., 1) in O(1) time by pointer manipulations. Since this paper, there was a question whether we can do the work with only arrays. There were two solutions by Takaoka [17] and Vajnovszki [20]. Both require O(kn) space apart from the main container of permutations, where k is the number of distinct items. Thus we have a question whether we can do the work with only O(n) extra space. Again we have two solutions by Korsh and LaFollette (2004) [11] and Takaoka and Violich (2005) [18]. Those algorithms are based on the two-level approach, which we use in this paper also. In [11], The upper structure is based on Johnson-Trotter algorithm and the lower structure is based on Eades and McKay [4] for combination generation. The upper structure for the latter method [18] is the same, Johnson-Trotter, and the lower is based on Chase's hard algorithm [3] for combination generation.

Suppose we have a multi-set $(1,…,1,2,…,2,……k,…,k)$, where there are $n[i]$ items of i, for i=1, …, k. The upper algorithm controls which item i to move, and the lower algorithm controls the movement of those n[i] identical items by regarding them as ones and others as zeros.

In this paper, we use Johnson-Trotter for the upper structure and Takaoka's algorithm [16] for combination generation for the lower structure, which is less restrictive than Eades-McKay and Chase. Both the upper structure and lower structure are based on the idea of tree traversal. Thus our algorithm has a more transparent and consistent design methodology and the program code is shorter than existing ones. To make the problem more visible we start from an example. The following is the list of permutations of (1, 2, 3, 4) by Johnson-Trotter, which should be read column-wise.

```
1 2 3 4      3 2 4 1      1 3 4 2      4 3 2 1      1 4 2 3      2 4 3 1
2 1 3 4      3 2 1 4      3 1 4 2      4 3 1 2      4 1 2 3      2 4 1 3
2 3 1 4      3 1 2 4      3 4 1 2      4 1 3 2      4 2 1 3      2 1 4 3
2 3 4 1      1 3 2 4      3 4 2 1      1 4 3 2      4 2 3 1      1 2 4 3
```

Table 1. Permutations of (1, 2, 3, 4) by Johnson-Trotter

In this list 1 moves right over (2, 3, 4). Then 2 moves right. Then 1 moves left over (3, 2, 4), etc., alternately. If we remove 1 at the top of each column, we have permutations (2, 3, 4), (3, 2, 4), (3, 4, 2), (4, 3, 2), (4, 2, 3), (2, 4, 3). This list of permutations has a similar pattern of the movement of 2 going back and forth alternately. Later we show an O(1) time algorithm for Johnson-Trotter.

The next list is the list of combinations of four items, called 4-combinations out of 6 items {1, 2, 3, 4, 5, 6} in in-place expression and binary vector form. The binary vector is for illustration purposes. In later sections, we generate only in-place forms. In combination generation, we generate combinations from ones at the left end to the right end, which we call the forward generation. When we use combination generation repeatedly as the lower structure of multiset permutation generation, we use forward generation and backward generation alternately. Backward generation is to generate combinations in reverse order of forward generation. We can repeat forward and backward with O(1) time re-initialization. We call this property "reversible". If we can repeat forward only with O(1) re-initialization, it is called repeatable.

```
         Forward Generation              Backward Generation
     In-place    binary vector        in-place     binary vector
    ------------------------------    --------------------------------
      1 2 3 4     1 1 1 1 0 0          4 3 6 5      0 0 1 1 1 1
      1 2 3 5     1 1 1 0 1 0          2 3 6 5      0 1 1 0 1 1
      1 2 3 6     1 1 1 0 0 1          2 3 4 5      0 1 1 1 1 0
      1 2 4 6     1 1 0 1 0 1          2 3 4 6      0 1 1 1 0 1
      1 2 4 5     1 1 0 1 1 0          2 5 4 6      0 1 0 1 1 1
      1 2 6 5     1 1 0 0 1 1          1 5 4 6      1 0 0 1 1 1
      1 3 6 5     1 0 1 0 1 1          1 3 4 6      1 0 1 1 0 1
      1 3 4 5     1 0 1 1 1 0          1 3 4 5      1 0 1 1 1 0
      1 3 4 6     1 0 1 1 0 1          1 3 6 5      1 0 1 0 1 1
      1 5 4 6     1 0 0 1 1 1          1 2 6 5      1 1 0 0 1 1
      2 5 4 6     0 1 0 1 1 1          1 2 4 5      1 1 0 1 1 0
      2 3 4 6     0 1 1 1 0 1          1 2 4 6      1 1 0 1 0 1
      2 3 4 5     0 1 1 1 1 0          1 2 3 6      1 1 1 0 0 1
      2 3 6 5     0 1 1 0 1 1          1 2 3 5      1 1 1 0 1 0
      4 3 6 5     0 0 1 1 1 1          1 2 3 4      1 1 1 1 0 0
```
Table 2. List of 4-combinations out of six items by Takaoka's algorithm [16]

We note that there is only one change from combination to combination in the in-place list, and in the binary vector list, a one moves over a block of ones, and does not change the relative order of zeroes. From 011110 to 011011, for example, the 4[th] element and 6[th] element are swapped, e. g., the 4[th] element, 1, goes over the 5[th] element, 1. In general, the 1 may go over several consecutive 1's in a larger example. Note also that four ones at the left end finally come to the right end of the binary vector, and vice versa for backward generation. We can modify the algorithm so that forward generation and backward generation alternate to fit into Johnson-Trotter.

Suppose we have a list of multiset, such as (1, 1, 2, 3, 3, 4, 4, 4), we move 1's to the right using Takaoka's algorithm by regarding them as ones and other items as zeroes. When all 1's arrive at the right end, that is, (2, 3, 3, 4, 4, 4, 1, 1), we perform one step of the movement of 2's to the right, and 1's start to move to the left, resulting in (1, 1, 3, 2, 3, 4, 4, 4). That is, the movement of each item in Johnson-Trotter is generalized by the movements of several identical items by a combination generation algorithm. In Korsh and LaFollette [11], the combination algorithm is that of Eades and McKay [4], and in Takaoka and Violich [18], it was Chase's [3] algorithm.

The following is the complete list of permutations of the multiset (1, 1, 2, 2, 3) by our algorithm.

| | | |
|---|---|---|
| 1 1 2 2 3 | 2 3 2 1 1 | 1 1 3 2 2 |
| 1 2 1 2 3 | 2 3 1 2 1 | 1 3 1 2 2 |
| 1 2 2 1 3 | 2 3 1 1 2 | 1 3 2 1 2 |
| 1 2 2 3 1 | 2 1 1 3 2 | 1 3 2 2 1 |
| 2 1 2 3 1 | 2 1 3 1 2 | 3 1 2 2 1 |
| 2 1 2 1 3 | 2 1 3 2 1 | 3 1 2 1 2 |
| 2 1 1 2 3 | 1 2 3 2 1 | 3 1 1 2 2 |
| 2 2 1 1 3 | 1 2 3 1 2 | 3 2 1 1 2 |
| 2 2 1 3 1 | 1 2 1 3 2 | 3 2 1 2 1 |
| 2 2 3 1 1 | 1 1 2 3 2 | 3 2 2 1 1 |

Table 3. List of permutations of (1, 1, 2, 2, 3)

**2. General Form of Loop-less Algorithms by Tree Traversal**

We borrow some materials from [16] to describe the concept of tree traversal used in this paper. Let $\Sigma = \{\sigma_0, \ldots, \sigma_{r-1}\}$ be an alphabet for combinatorial objects. A combinatorial object is a string $a_1\ldots a_n$ of length n such that each $a_i$ is taken from $\Sigma$ and satisfies some property. An order is defined on $\Sigma$ with $\sigma_i < \sigma_{i+1}$. Let $\Sigma^n$ be the set of all possible strings on $\Sigma$ of length n with the lexicographic order <. The order < on a set of combinatorial objects, $S \subseteq \Sigma^n$, is defined by projecting the lexicographic order on $\Sigma^n$ onto S.

The lexicographic tree, or lexico-tree for short, of S is defined in the following way. Each **a** ∈ S corresponds to a path from the root to a leaf. The root is at level 0. If **a** = $a_1\ldots a_n$, $a_i$ corresponds to a node at level i. We refer to $a_i$ as the label for the node. We sometimes do not distinguish between node and label. If **a** and **b** share the same prefix of length k, they share the path of length k in the tree. The children of each node are ordered according to the labels of the children. A path from the root to a leaf corresponds to a leaf itself, so **a** corresponds to the leaf. The combinatorial objects at the leaves are thus ordered in lexicographic order on S.

The twisted lexico-tree of a set S of combinatorial objects is defined as follows together with the parity function. We proceed to twist a given lexico-tree from the root to leaves in a breadth-first search manner. Let the parity of the root be even. Suppose we processed up to the i-th level. If the parity of a node v at level i is even, we do not twist the branches from v to its children. If the parity of v is odd, we arrange the children of v in reverse order. If we process all nodes at level i, we give parity to all the nodes at level i+1 from first to last alternately starting from even. We denote the parity of node v by parity(v). When we process nodes at level i in the following algorithms, which are children of a node v such that parity(v)=p, we say the current parity of level i is p. Note that (labels of) nodes at level i are in increasing order if the parity of the parent if even, or equivalently if the current parity of level i is even. If the parity is odd, they are in decreasing order. We draw trees lying horizontally for notational convenience. We refer to the top child of a node as the first child and the bottom as the last child. Note that the parity is defined globally on the entire tree. Other papers define the parity locally on the children of each node.

If the labels on the paths from the root to two adjacent leaves in the twisted lexico-tree for S are different at a fixed number of nodes, we can generate S from object to object with the fixed number of changes. We design an efficient algorithm that traverses the twisted lexico-tree and generate combinatorial objects in O(1) time per object. Thus the fixed number of changes is a necessary condition for our algorithm, but not sufficient.

The current parity at level i is given by parity[i], and the procedure "output" is only to see the result; for O(1) time this will be eliminated. We allow O(n) time for initialization, but for the repeated use of generation cannot afford to re-initialize arrays in O(n) time. This will be addressed in later sections. The parity 0 is for even and 1 for odd. Let $\Sigma = \{0, ... , r-1\}$. We associate an object with a leaf of a tree. The path from the root to the leaf identifies the object. Traversal from a node to the next sibling node corresponds to generating the next object, making some changes on the object. We maintain the parity information for each level in array "parity". The level of the tree corresponds to various aspects of the objects as we will see in later sections. A generic form of the algorithm follows. All algorithms are given in C or pseudo code.

**Algorithm 1.** Iterative tree traversal in pseudo code. i keeps track of the current level of the tree.

1. initialize array a to be the first object in S;
2. initialize $v_1, ... , v_n$ to be nodes on the path to the first object (top path);
3. **for** i:=0 **to** n **do** up[i]:=i;  // up[i] shows where to jump up from level i
4. **for** i:=0 **to** n **do** parity[i]:=0;
5. **repeat**
6.    **output**(a);
7.    i:=up[n]; up[n]:=n;
8.    perform changes on a at $v_i$ and related positions;
9.    let $v_i$ go to the next node at level i based on the current parity;
10.    **if** $v_i$ is the last child of its parent **then begin**
11.      let $v_i$ further go to the next node at level i;
12.      up[i]:=up[i-1]; up[i-1]:=i-1;
13.      parity[i]:=1-parity[i]
14.    **end**
15. **until** i=0.

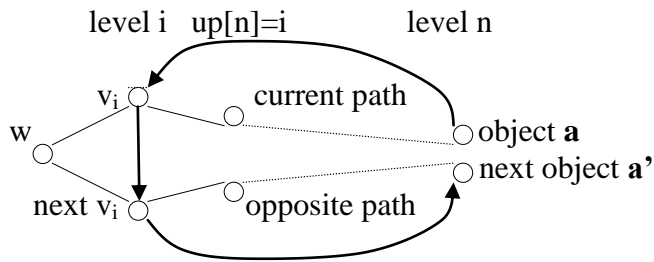

When we come to the last child of a parent (w in the above figure), we have to update up[i] to up[i-1] so that when we visit the last leaf of the sub-tree rooted at w, we can come back directly from the leaf to w or its ancestor if w itself is a last child. We refer to the paths from $v_i$ to **a** and from next $v_i$ to **a'** as the current path and the opposite path. A current path and opposite path consist of last children and first children respectively except for the left ends. If next $v_i$ in the above figure is a last child, we further set $v_i$ to the next node of next $v_i$, say u, so we can avoid O(n) time to set up the environment for such u's later. This is illustrated in the above figure by the path from "next $v_i$" to "next object **a'** ". That is, when we traverse down the current path, we prepare for the opposite path so that we can jump over the opposite path from level i to the leaf. In this situation **a** and **a'** share the same prefix from position 1 to i-1. We call i the difference point. The two strings can also share the same suffix (empty in the above figure). Let the longest such is from position j+1 to n. Then we call j the solution point. Intuitively this is the point where the difference caused upstream is solved.

**3. Review of Johnson-Trotter**
In the iterative algorithm for Johnson-Trotter, the level of the tree corresponds to the item we are trying to move. Parity corresponds to the direction of movement, left or right, represented by -1 (odd parity) or +1 (even parity). Array "p" is to hold the position of item x in array "a". Array element "c[i]" is to count the number of child nodes at level i and check the last child condition. Procedure "move(x)" is to move item x to the direction given by d[x], and update the positions of affected items. The comment /*output */ shows the point where a new permutation is generated. Array "d" plays the role of parity in Algorithm 1. In the program, level variable i corresponds to item i. If we regard the position of each item as the label of the node, we have a close correspondence with the twisted lexico-tree.

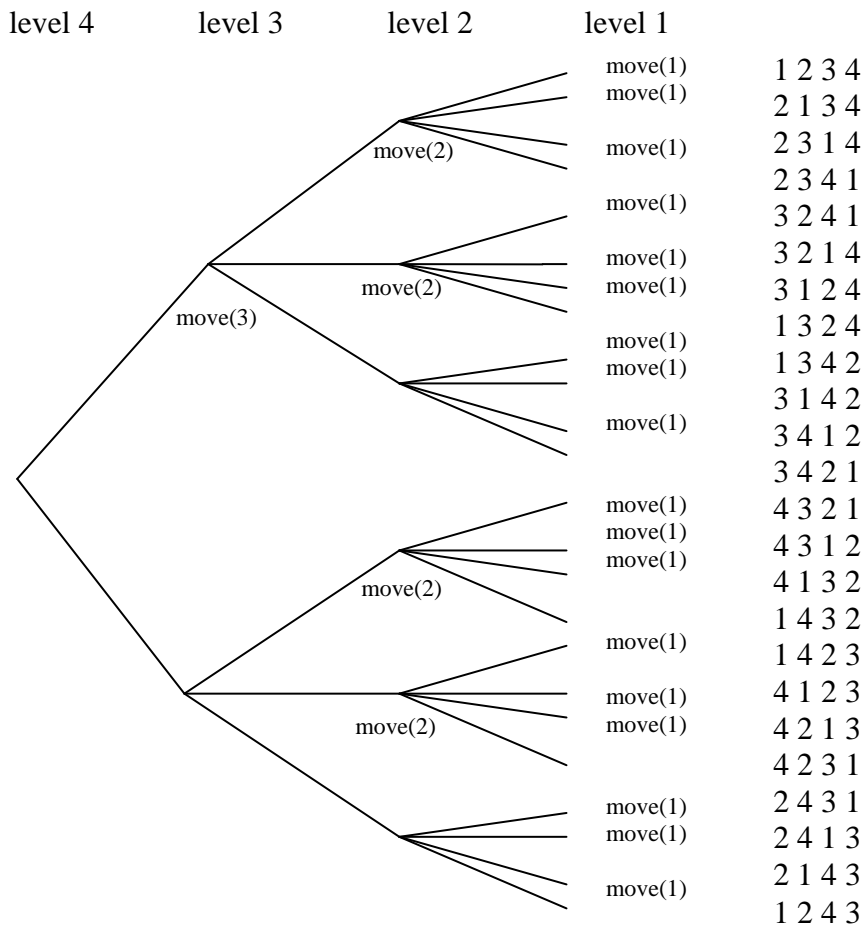

Figure 1 Tree for Johnson-Trotter

In this figure levels increase from leaves to the root. Variable i keeps track of the current level of tree traversal. The action move(x) is attached to each node of the tree in the figure. Variable $c[i]$ is called the child counter, which can tell us if we are hitting a last child at the current level. In the following sections, we say x moves actively in the procedure call "move(x)", and the item neighboring x moves passively by swapping.

**Algorithm 2**. Iterative algorithm for Johnson-Trotter
**procedure** move(x);
**begin** var w;
   w:=a[p[x]+d[x]]; a[p[x]+d[x]]:=x; a[p[x]]:=w;
   p[w]:=p[x]; p[x]:=p[x]+d[x]; c[i]:=c[i]+1; /* output */
**end**;
**begin** {main program}
  **for** i:=1 **to** n **do begin**
    a[i]:=i, up[i]:=i; p[i]:=i; c[i]:=1; d[i]:=1;
  **end**;
  **repeat**
    i:=up[1]; up[1]:=1;
    move(i);
    **if** c[i] = n-i+1 **then begin** /* currently hitting a last child at level i */
      up[i]:=up[i+1]; up[i]:=i+1; /* inherit the up-value of the parent */
      c[i]:=1;
      d[i]:=-d[i];                 /* parity (direction) reversed */
    **end**
  **until** i=n;
**end**..

## 4. The Two Level Approach

Now we combine the loop-less algorithm in Section 3 for permutations and the combination generation algorithm described in the Appendix 1 in detail. Let us generate permutations of multiset (1,1,…,1,2,2,…,2,…,k,k,…,k). Let n[i] be the number of items i in the multiset, that is, the multiplicity of i. Let R=n[1]+ … + n[k], that is, the total size of the multiset.

     When items 1 move in the container M[1 .. R], the range of movement is [1 .. R], the whole span of array M. The range of item 2 is limited to [1 .. R-n[1]] when all 1s are at the right end, or range [n[1]+1 .. R] if all 1s are at the left end. If we define the range of 2 as [1 .. R-n[1]], the latter range can be obtained by adding the base, base[2]=n[1]. Note that when we talk about movement, we mean active movement, not passive movement which is done by lower numbered items. Thus items k never moves actively. This idea leads to the definition of limit[i] and base[i] for item i. The value of limit[i] is defined to be the upper bound of the above described range.

    base[0]=0, base[i] = n[1] + … + n[i-1], i=1, …, k-1

    limit[i] = R – base[i], i=1, …, k-1

   We maintain positions of items i by the combination generation algorithm. Suppose the positions of items i changed from (q[1], …, q[j], …, q[n[i]]) to (q[1], …, q'[j], …, q[n[i]]) by the combination generation algorithm. Then we see that an item i moves from position q[j] to position q'[j] in the range of item i in the main container array M. In Algorithm 4, those values of q[j] and q'[j] are returned by the call to "combination-server" in global variables "from" and "to". When d[i-1]>0, items (i-1) are ready to move right at the left end at the next step, the positions for the movement of item i need to be adjusted by adding base[i] to the positions in the range. Let C(n[i], R) be the binomial coefficient of n[i]-combinations out of R items for the last child check.
**Example**. Suppose the first item of 2 moved right from (1,1,1,2,3,3,2) to (1,1,1,3,2,3,2). As 2's are moving right globally, the positions of 2's before and after the move are (4, 7) = 3+ (1, 4) and (5, 7)=3+(2, 4). This is because we add the base 3 to 2-combinations of items 2 in the in-place notation, (1,4) and (2,4) out of 4 items {1, .., 4}. Thus 2 moves from position 4 to 5.

In the following, the work "maintain combinations" includes "return the next combination".

**Algorithm 3**. Multiset Permutation Generation

Lower Level
Combination-server   /* This server performs the following operations on demand */
**begin**
   Initialize n[1]-combinations out of limit[1] items
   Initialize n[2]-combinations out of limit[2] items
   …
   Initialize n[k-1]-combinations out of limit[k-1] items

   Maintain n[1]-combinations out of limit[1] items
   Maintain n[2]-combinations out of limit[2] items
   …
   Maintain n[k-1]-combinations out of limit[k-1] items
**end**

Upper Level
**procedure** move(x);
**begin** var w;
   c[i]:=c[i]+1;
   Let the change of combination be from "from" to "to";
   if d[i-1]>0 then from=from + base[i]; to:=to + base[i];
   w:=a[from]; a[from]:=a[to]; a[to]:=w;  /* output */
**end**;
**begin** {main program}
   **for** i:=1 **to** k-1 **do** call combination-server to
                     initialize n[i]-combinations out of limit[i] items;
   **for** i:=1 **to** k **do begin**
      a[i]:=i, up[i]:=i; p[i]:=i; c[i]:=1; d[i]:=1;
   **end**;
   **repeat**
      i:=up[1]; up[1]:=1;
     call combination-server for next combination for item i;
      **if** i<k **then** move(i);
      **if** c[i]  = C(n[i], R) **then begin** /* checking a last child */
         Re-initialize combinations of item i for reverse direction;
         up[i]:=up[i+1]; up[i]:=i+1;
         c[i]:=1;
         d[i]:= −d[i];
      **end**
   **until** i=k;
**end**.

Re-initialization of combinations of item i must be done in O(1) time. This will be mentioned in Appendix 1. The value of C(n[i], R) is potentially large, and may not be contained in a single precision variable. We can use the termination condition (i=0) in the Algorithm 4 in Appendix 1, and bring the effect to the calling site. As initialization for combinations of item i takes O(n[i]) time, Algorithm 3 takes O(R) time at the beginning. As the size of an object of n[i]-combinations is O(n[i]), the total space requirement is O(R). In the Appendix 2 we declare fixed-sized arrays for readability. It is straightforward to allocate necessary space dynamically using the "calloc" function in C.

Takaoka's algorithm for combination generation is designed by tree traversal, and has the following desirable property that fits into the upper algorithm of Johnson-Trotter as the following fact shows. The details are given in Appendices.

**Fact.** Takaoka's algorithm generates combinations of k items out of n items in O(1) time per combination in O(k) space. In the hypothetical binary vector of size n for a combination, the movement of 1 goes over zero or more consecutive 1s. The relative order of 0s is not changed. In the binary vector, 1s start from the left end and end at the right end. If we call the range [1, .., limit[i]] the capsule of items i, the relative order of items i in the capsule is not changed by the movement of lower items. With O(1) time re-initialization, the algorithm continues to generate the same set of combinations in reverse order.

**5**. **O(1) time generation of the parking function – A three level design**
The parking function [9] is a set of integers that assigns cars to parking lots and satisfy some conditions.

Cars and lots are numbered 1 to n. The parking place has a one-way allay, and car i is given a parking lot p(i) for each i. Lots are numbered consecutively along the allay. If car i finds lot p(i) is occupied it can go to a lot whose number is greater. Every car needs to be parked. The mapping p is called a parking function. The problem is to generate all possible parking functions in O(1) time per function.

**Example.** Let n=3. (1,2,2) is a solution. Car 1 is given lot 1. Cars 2 and 3 are given lot 2. If car 2 or 3 finds lot 2 is occupied, it can go to lot 3. As the assignment can permute, (2,1,2) is also a solution. This is a multiset permutation. How about (1,3,3)? If car 2 occupies lot 3, car 3 gets trouble, and vice versa.

Let us call a solution standard if $p(1) \leqq \cdots \leqq p(n)$ is satisfied. The standard solutions are the Catalan sequences, which further satisfy $p(i) \leqq i$. In [16], we have an O(1) time algorithm for the Catalan sequences, which we call "Catalan". Here we propose to use Catalan as the upper algorithm and the multiset permutation generation algorithm, called "Multiperm", in the previous section as the lower structure. The combination mechanism for those two is of type 1 in Section 2, resulting in three levels in total. How to combine those two needs some care.

If not impossible, Multiperm seems to be difficult to run repeatedly or reversely with O(1) re-initialization. Thus we propose to use what is called "time stealing" in [18]. This is to prepare two identical copies of Multiperm together with data structures, named Multiperm1 and Multiperm2. As the size of the set of multiset permutations is greater than the size of the parking function, n, it is easy to re-initialize Multiperm2 while Multiperm1 is running, and vice versa, within the constraint of O(1) time per parking function.

Thus we claim that we can generate parking functions in O(1) time with O(n) space.

## 6. Concluding Remarks
The cpu times in seconds of the three algorithms are given in the following table.

| n[1], …, n[5] | Korsh & LaFollette | Takaoka & Violich | This paper |
|---|---|---|---|
| 3 3 3 3 3 | 52.1 | 28.2 | 26.2 |

CPU Intel(R) Pentium(R) M Processor 1300MH, gcc compiler

We showed how to design a loop-less algorithm for multiset permutation generation based on a two-level approach. To avoid complications, we used an abstract algorithm, called "combination server", which delivers combinations of various sizes one by one at the request of the upper algorithm of Johnson-Trotter. If we implement our algorithm completely in a procedural language, we can use two-dimensional arrays, another dimension corresponding to which kind of items are moving. The source code in Appendix 2 is based on this approach by two dimensional arrays. Note that with this approach, memory requirement can be O(R), where R is the total size of each permutation. Precisely speaking, we would need to maintain eight arrays of size n[i] for each item i. Those arrays need to be maintained by pointers for dynamic memory allocation, although the main container of permutations remains to be a fixed array of size R. In the upper structure of the algorithm we need seven arrays of size k where k is the number of distinct items. In [11], 20 arrays of various sizes are used and in [18] an array of size R and nine arrays of size k are used. It remains to be seen whether we can further simplify the algorithm or reduce memory requirement. Note that the repeatability or reversibility of the lower level combinatorial generation with O(1) re-initialization is essential. We used the latter for our multiset permutation generation. It is interesting to see if the former will work in our case. It works for parenthesis strings as shown in Takaoka and Violich [18 ].

## Appendix 1. Review of Takaoka's algorithm for in-place combination generation.

We describe an O(1) time algorithm for generating the set S of combinations of n elements out of the set {1, ..., r}. In the following array "q" is the main container of combinations and "a" is an auxiliary array for book-keeping; it keeps track of tree traversal with some delay. We state a few lemmas describing necessary properties for our multi-set permutation generation.

**Example** . The twisted lexico-tree for combinations of 4 elements out of 6 elements is given with the contents of array "a" and "q" at the leaves. A white circle is for the even parity and black for odd.

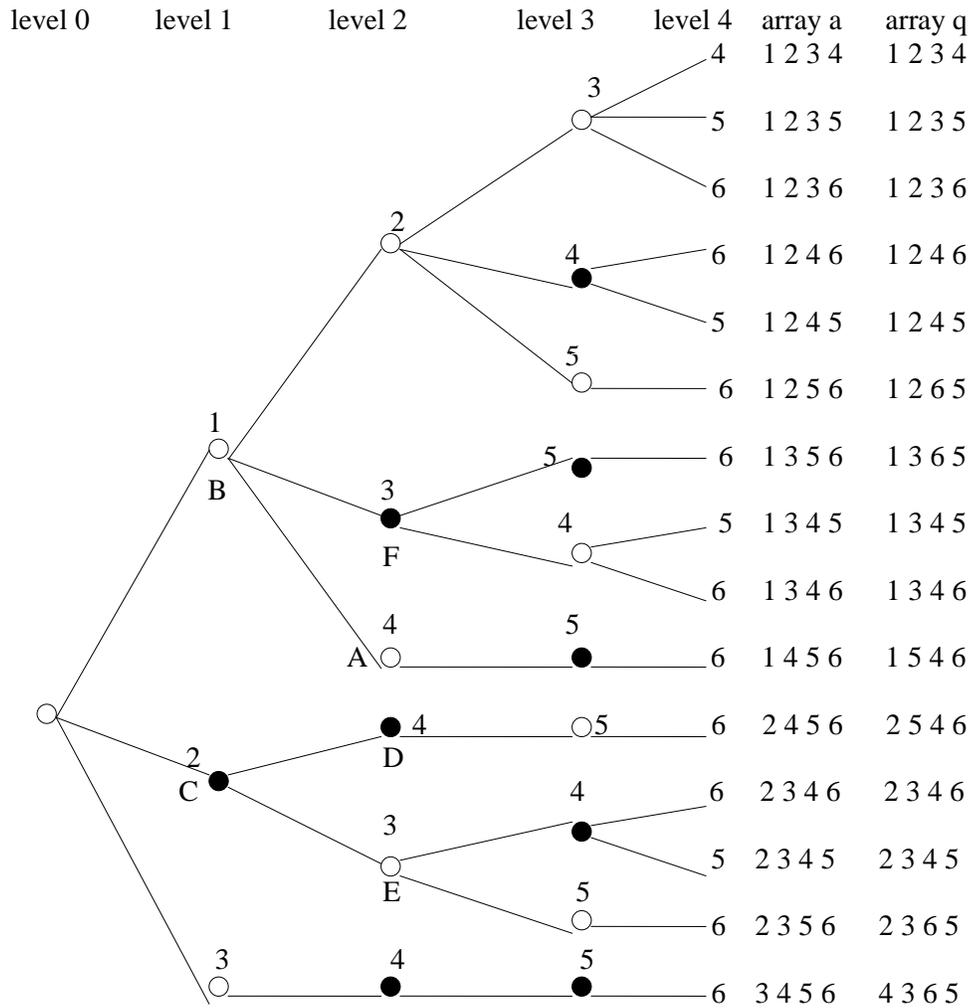

**Lemma 1**. The set S generated by the twisted lexico-tree generates combinations with one change per combination. See [16] for a proof.

   Now we describe implementation details. There are many nodes which have only one child down to the leaf, causing straight lines. Traversal of these lines downwards will cause O(n) time. Having one child is caused by a node with the maximum possible label at the level. To control the position to which we come down, we keep an array "down". In the above example, we use "up" to go from A to B, and use "down" to go from C to D. A snapshot of the movement is like (..., F,A,B,C,D,E, ...). As we see below, we cannot generate combinations in labels attached to the nodes of the twisted lexico-tree with a fixed number of changes. Let array **a** contain those labels. Then we have a situation with

$$\mathbf{a} = (a_1, ... ,a_{i-1}, a_i, ... ,a_j, ... ,a_n) \text{ and } \mathbf{a'} = (a_1, ... ,a_{i-1}, a_i+1, ... ,a_j+1 ... ,a_n),$$

where $a_{k+1}=a_k+1$ for $i \leq k < j$, or a symmetric case where $\mathbf{a'} = (a_1, ... ,a_{i-1}, a_i-1, ... ,a_j-1, ... ,a_n)$, where $a_{k+1}=a_k-1$. Note that "i" is the difference point and "j" is the solution point. In the first case, "$a_i$" goes out of the combination and "$a_j+1$" comes into the combination like a revolving door system. We keep the combinations in array "q" and use array "a" for book-keeping. As we cannot change j-i+1 places in "a" in O(1) time, we just change $a_i$ and $a_j$, and change the contents of q correspondingly. Thus we do not implement line 11 of Algorithm 1 to prepare for the opposite path to the next object. To keep track of the positions of these elements in q, we use array "pos". Solution points are maintained in array "solve". Since some parts of array "a" are not maintained to the corresponding labels in the tree, we use Boolean array "mark" to show no maintenance. When mark[i]=true, the proper value of a[i-1] is given by its child a[i]. The main part of the algorithm follows in which up[i], down[i], solve[i], and pos[i] are initialized to i for all i. The values of d[i] are initialized to 1 and those of "mark" to false. Line-by-line explanation follows the algorithm. If we change the termination condition at line 23 to **false** we can keep generating combinations in forward order and reverse order alternately forever. The proof of the next lemma is omitted.

**Lemma 2**. The generation of combinations in the binary vector form by the above twisted lexico-tree does not change the relative order of 0's. Items 1 move from the left end and finish at the right end for one run in the binary vector. From the second run on, the movement alternates in direction.

To continue to generate combinations in reverse order, we can simply perform i:=down[i], and repeat from line 3. In the source list in the Appendix 2, we use Algorithm 4 as "combination-server", where all variables are indexed by item. Thus simple variables become one-dimensional arrays and one-dimensional arrays become two-dimensional arrays. Re-initialization in Algorithm 3 can be implemented by performing i[I]=down[I][i[I]] for item I. Also checking the last child for the combination server can be implemented by testing "i[I]=0". In the program, Algorithm 3 and Algorithm 4 communicate through global variables, which make the algorithm structure less transparent, but efficiency is gained. When Algorithm 4 is used as "combination-server", line 2 and 23 will be removed.

**Algorithm 4**. In-place algorithm for combinations {arrays a, q, d, up, pos, sol, solve, mark, used}
1. d[n+1]:= -1; up[0]:=0; i:=n; initialize d[i] to 1, and other arrays to i for all i; d[0]:=0;
2. **repeat**
3.    **output**(q);   /* This is to output q */
4.    **if** mark[i]=**true then begin** a[i-1]:=a[i]-1; mark[i]:=**false end**;
5.    **if** d[i+1]<0 **then begin**   /* Moving from an even node to an odd node*/
6.       q[pos[a[i]]]:=a[i]+d[i]; pos[a[i]+d[i]]:=pos[a[i]]
7.    **end else begin**
8.       **if** d[i]>0 **then**    /* Moving from odd to even, with "a" value increasing */
9.          q[pos[a[i]]]:=a[solve[i]]+d[i]; pos[a[solve[i]]+d[i]]:=pos[a[i]]
10.    **end else if**  d[i]<0 **then begin**   /* Moving from odd to even, with "a" value decreasing */
11.       q[pos[a[solve[i]]]]:=a[i]+d[i]; pos[a[i]+d[i]]:=pos[a[solve[i]]]
12.    **end**;
13.    a[i]:=a[i]+d[i];
14.    **if** d[i+1]>0 **then** a[solve[i]]:=a[solve[i]]+d[i];

15.   up[i]:=i;
16.   **if** (d[i]>0) **and** (a[i]=r-n+i) **or** (d[i]<0) **and** (a[i]=a[i-1]+1) **then begin**
17.      up[i]:=up[i-1]; up[i-1]:=i-1;
18.      down[up[i]]:=i;
19.      **if** d[i]<0 **then begin** solve[[up[i]]:=i; mark[i]:=**true end**;
20.      d[i]:= -d[i];
21.      **if** (d[i]<0) **or** (i=n) **then** i:=up[i] **else** i:=down[i]
22.   **end else** i:=down[i]
23. **until** i=0;  { This termination condition is replaced by **false** for an infinite loop}
24. **output**(q).

Line 4: Update the parent when maintenance is not done.
Lines 5-12: Update the contents of q.
Lines 5-7: Note that d[i+1] was altered in a previous step. Thus we regard parity($a_i$) as even despite d[i+1]<0. Move from B to C with i=1 for example in the figure.
Lines 8-10: Similarly move from F to A with i=2 for example.
Lines 10-12: Similarly move from D to E with i=2 for example.
Line 13: Take the next child.
Line 14: Change the value of a at the solution point.
Line 15: Originate the destination for descendants to come up to. This value will possibly propagate down in line 17 later.
Line 16: Check if the node is a last child.
Line 17: Compute the correct value of "up".
Line 18: Let the ancestor know where to come down next.
Line 19: If parity is odd, update the solution point for ancestor by i. This may further be updated by larger i. Also signal that the parent a[i-1] is not updated for the next object. If up[i]=i-1, a[i-1] has been given a proper value of a[i]-1, but signaling "mark[i]=true" will not cause any harm.
Line 20: Change the parity at i.
Line 21: Go up if you hit a last child with even parity or at a leaf. Otherwise go down.
Line 22: Go down if you do not hit a last child.
Line 23: Go out of iteration if you come to the root, or go back to line 2 for repeated generation.

**Appendix 2. Program Source List for Multiset Permutations in C**

For the readability this program is based on two-dimensional arrays for local variables for combination generation. Available at http://www.cosc.canterbury.ac.nz/tad.takaoka/multiperm.c

```
// This program generates multiset permutations of at most 10 kinds of items of total
//  size 20. Upper-case is for upper-level objects, lower-case for lower-level objects
int tt, I, K, N[10], A[10], P[10], D[10], UP[10]; // A and P are not essential
int R, count; int countk, init, COUNT, from, to;
int M[40], base[10], limit[10];
int a[10][20], q[10][20], up[10][20], solve[10][20],
    d[10][20], pos[10][20], mark[10][20], down[10][20],
    i[10];
int true=1, false=0;
init_base(){int i, k;
  base[1]=0;
  for(k=2;k<=K;k++)base[k]=base[k-1]+N[k-1];
  for(k=1;k<=K;k++)limit[k]=R-base[k];   // limit[k] size of the world for item k
```

```
}
OUT(){int i;
   for(i=1;i<=R;i++)printf("%d ",M[i]);
   printf("COUNT=%d\n", COUNT);
}
MOVE(){int w;
   if(D[I-1]>0){from=from+base[I]; to=to+base[I];}
   w=M[from]; M[from]=M[to]; M[to]=w;         // 3 3 2 2 1
   COUNT++;  OUT();                            //   *     base[2] =2
}
main(){int k, j;
printf("Input K :number of kinds of items ");
scanf("%d", &K); getchar(); // no of distinct items
printf("Input multiplicities of each kind of items ");
for(I=1;I<=K;I++)scanf("%d",&N[I]);
getchar();
R=0; for(I=1;I<=K;I++) R=R+N[I]; // Total size
// M is the main container of permutations. M is initialized to (1,1,1,2,2,3,3...) etc.
tt=clock();
init=1;
for(countk=1;countk<=K;countk++){
   for(I=init;I<=init+N[countk]-1;I++){
      M[I]=countk;
   }
   init=init+N[countk];
}
printf("M= ");
for(I=1;I<=R;I++)printf("%d ", M[I]); printf("\n");
init_base();
for(I=1;I<=K-1;I++){combi_init(I, N[I], limit[I]);}
/*** Johnson-Trotter for upper structure ***/
for(I=1;I<=K;I++){A[I]=I; D[I]=1; P[I]=I; UP[I]=I;}
COUNT=1; OUT();
do{
   I=UP[1]; UP[1]=1;
   combi(I, N[I], limit[I]);
   if(I<K)MOVE(I);
   if(i[I]==0){
      i[I]=down[I][i[I]];
      UP[I]=UP[I+1]; UP[I+1]=I+1;
      D[I]=-D[I];
   }
}while(I!=K);
printf("time %d\n", clock()-tt);
}
/*** Initialization for combination server ***/
combi_init(int t, int n, int r){
   { int i;
   for(i=1; i<=n; i++){a[t][i]=i; q[t][i]=i; d[t][i]=1;}
```

```
    for(i=0; i<=n; i++){up[t][i]=i; down[t][i]=i;
            pos[t][i]=i; solve[t][i]=i;}
  }
  d[t][n+1]= -1; up[t][0]=0; i[t]=n;
}
/*** Combination server for lower structure ***/
combi(int t, int n, int r){ // t for item t
    int ii; // ii is level indicator
    ii=i[t];
    if(mark[t][ii]==true)
      {a[t][ii-1]=a[t][ii]-1;
      mark[t][ii]=false ;}
    if(d[t][ii+1]<0) {
      from=q[t][pos[t][a[t][ii]]];
      q[t][pos[t][a[t][ii]]]
        =a[t][ii]+d[t][ii];
      to=q[t][pos[t][a[t][ii]]];
      pos[t][a[t][ii]+d[t][ii]]=pos[t][a[t][ii]];
    } else
    if(d[t][ii]>0) {
      from=q[t][pos[t][a[t][ii]]];
      q[t][pos[t][a[t][ii]]]=a[t][solve[t][ii]]+d[t][ii];
      to=q[t][pos[t][a[t][ii]]];
      pos[t][a[t][solve[t][ii]]+d[t][ii]]=
        pos[t][a[t][ii]];
    } else if(ii>0){
      from=q[t][pos[t][a[t][solve[t][ii]]]];
      q[t][pos[t][a[t][solve[t][ii]]]]=a[t][ii]+d[t][ii];
      to=q[t][pos[t][a[t][solve[t][ii]]]];
      pos[t][a[t][ii]+d[t][ii]]=pos[t][a[t][solve[t][ii]]];
    }
    a[t][ii]=a[t][ii]+d[t][ii];
    if(d[t][ii+1]>0) a[t][solve[t][ii]]=a[t][solve[t][ii]]+d[t][ii];
    up[t][ii]=ii;
    if(((d[t][ii]>0)&&(a[t][ii]==r-n+ii))||
        ((d[t][ii]<0)&&(a[t][ii]==a[t][ii-1]+1))){
      up[t][ii]=
        up[t][ii-1]; up[t][ii-1]=ii-1;
      down[t][up[t][ii]]=ii;
      if(d[t][ii]<0) {solve[t][up[t][ii]]=ii;
        mark[t][ii]=true;}
      d[t][ii]= -d[t][ii];
      if((d[t][ii]<0) || (ii==n))
        i[t]=up[t][ii]; else i[t]=down[t][ii];
    } else i[t]=down[t][ii];
```